\documentclass[epsf,preprint,aps,12pt,showpacs,nofootinbib,tightenlines]{revtex4}

\usepackage[section]{placeins}
\usepackage{graphicx}
\usepackage{dcolumn}
\usepackage{bm}

\usepackage{epsfig}
\usepackage{amssymb}
\usepackage{amsmath}
\usepackage{flafter}
\usepackage{array}

\usepackage[dvips,usenames]{color}

\newlength{\dinwidth}
\newlength{\dinmargin}
\begin{document}
\title{Nonrelativistic Cousin of QCD}
\author{Feng Wu}
\email[Electronic address: ]{fengwu@ncu.edu.cn}
\affiliation{%
Department of Physics, Nanchang University,
330031, China}

\begin{abstract}
Based on the uniqueness and universality of gravity, it is clear that theories with different dynamical exponents are related in the holographic approach. Concretely, we construct an M-theory background from pure QCD dual and show that a deformed $Sch_{6}^{4}$ geometry is obtained by compactification from the same background. The deformed $Sch_{6}^{4}$ geometry is considered as the geometrical realization of a four-dimensional nonrelativistic field theory. Several aspects of this nonrelativistic field theory are studied in the holographic picture. 
\end{abstract}
\maketitle 
\newpage

\date{\today}

\section{Introduction}
One of the most remarkable things we know about M-theory is the gauge/gravity duality. It provides a powerful tool which maps holographically a strongly interacting field theory to a weakly coupled gravitational theory so that one is able to calculate correlation functions in a strongly interacting system pretty much by hand. Regarding AdS/CFT correspondence \cite{Maldacena, Witten1, Gubser} as a paradigm and inspired by it, lots of research have been stimulated on understanding the strongly coupled physical systems through this duality.

In real world, strongly coupled systems appear in both relativistic and nonrelativistic physics. For the relativistic case, the most recognised example is strong interactions, which is described by Quantum Chromodynamics (QCD) in the infrared region. Currently, one of the most reliable holographic models QCD is the Sakai-Sugimoto model \cite{Sakai}. It consists of the "probe" $N_{f}$-folded flavor $D8$-$\bar{D8}$ branes in a classical supergravity background of the $N_{c}$-folded color $D4$ branes wrapped on a circle in the type IIA superstring theory. This particular  $N_{c}$-folded $D4$-brane configuration was proposed by Witten \cite{Witten2} as a holographic description of pure QCD. Sakai-Sugimoto model possesses QCD features such as chiral symmetry breaking and color confinement, and allows one to calculate the low energy meson spectrum.

In the nonrelativistic case, recently holographic descriptions of fermions at universality \cite{fermions} have been explored \cite{Son, Balasubramanian}. It is based on the speculation that for a nonrelativistic scale invariant field theory there exists a gravitational dual. In \cite{Kachru}, the formulation of dual gravitational models of theories with non-trivial dynamical exponents was constructed. Thus using holographic principle to study strongly coupled interacting theories has been extended to condensed matter systems near a critical point. Related topics have been studied in \cite{others}.

One of the most significant features of the holographic idea is scaling invariance. Though not yet being fully understood, according to the holographic approach, both relativistic and nonrelativistic field theories have gravitational models as their holographic duals. Since gravitational interactions are unique and universal, it is natural to speculate that every relativistic field theory is dual to a different nonrelativistic field theory and vise versa. Note that by this we mean the correspondence between two different theories rather than taking different energy limits of the same theory.

In this paper, we will explore this idea. We focus concretely on holographic pure QCD and find the nonrelativistic field theory to which it corresponds to. More precisely, we will perform a Null Melvin Twist \cite{NMT1, NMT2} to the background of type IIA $D4$ branes wrapped on a circle, show that the resulting geometry can be realized in the M-theory context, and identify that it is a four-dimensional field theory with dynamical exponent $z=4$ in the UV to which this geometry is dual. The techniques we used can certainly be applied to other systems.

For early work on the Null Melvin Twist to Sakai-Sugimoto model, see \cite{Pal}. Different from \cite{Pal}, we perform the Null Melvin Twist along the $D4$-brane world-volume directions instead of the transverse compact directions. Embedding nonrelativistic geometries into string theory or M-theory is previously studied in \cite{Herzog, Oz}.

\section{The pure QCD dual}
We begin with a brief review of the holographic pure QCD model \cite{Witten2}. It consists of $N_{c}$ $D4$ branes wrapped on a circle with anti-periodic boundary condition imposed for fermions. The near horizon limit of type IIA supergravity solution of $N_{c}$-folded $D4$ branes compactified to a $S^{1}$ circle in the $\tau$ direction is given by 
\begin{eqnarray}
ds^2&=&\left({U\over U_{R}}\right)^{{3\over 2}} \left( \eta_{\mu\nu} d x^{\mu} x^{\nu} + f(U) d \tau^2 \right)+ \left({U\over U_{R}}\right)^{-{3\over 2}} \left( {dU^2 \over f(U)} + U^2 d \Omega_{4}^{2} \right), \nonumber\\
e^{\phi} &=&g_{s} \left({U\over U_{R}}\right)^{{3\over 4}}, \;\;  F_{(4)} = {2\pi N_{c} \over V_{4}} \epsilon_{4},  \;\; f(U)= 1- \left({U_{KK} \over U} \right)^{3} 
\label{d4}
\end{eqnarray}
where $ \eta_{\mu\nu} d x^{\mu} x^{\nu}$ is the four-dimensional Lorentz metric, and $U$ and $\Omega_{4}$ are the five-dimensional Euclidean polar coordinates. $\epsilon_{4}$ is a volume form of $S^4$ whose volume is $V_{4}={8\pi^2 \over 3}$. $\phi$ is the dilaton and $F_{(4)}$ is the four-form Ramond-Ramond field strength. The constant $U_{R}$ is given in terms of the string coupling constant $g_{s}$ and the string length $l_{s}$ by 
\begin{eqnarray}
U_{R}^{3}=\pi g_{s} N_{c} l_{s}^{3}
\label{R}
\end{eqnarray} 
To avoid a conical singularity, the period of $\tau$ is determind to be 
\begin{eqnarray}
\delta\tau= {4\pi \over 3} \left( {U_{R}^{3}\over U_{KK}} \right)^{{1\over2}}.
\label{period}
\end{eqnarray} 

The symmetry group of the metric (\ref{d4}) is $ISO(1,3)\otimes SO(5)$ where $ISO(1,3)$ is the Poincare group of the $D4$-brane world-volume and $SO(5)$ is generated in the five transverse directions. From the four-dimensional perspective, all the fermions are massive with mass of order $\delta\tau^{-1}$ because of the anti-periodic boundary condition. Without symmetry protection, scalar fields would acquire mass of the same order through loop corrections. Thus, supersymmetry is completely broken and the only massless fluctuation modes of open strings attached to the $D4$ branes are Yang-Mills fields $A_{\mu}$. From the expansion of the DBI action in powers of the field strength 
\begin{eqnarray}
S_{D4}&=&T_{D4} \int d^4 x d \tau e^{-\phi} Tr \left( \sqrt{-det(\eta_{AB}+2\pi l_{s}^{2} F_{AB})} \right) \nonumber\\
      &\sim & T_{D4} {(2\pi l_{s}^{2})^{2} \delta\tau \over 2 g_{s}} \int d^4x \; Tr F_{\mu\nu}^{2} + O(F^{4}) 
\label{DBI}
\end{eqnarray} 
with the zero-point energy neglected and the $D4$-brane tension $T_{D4}$ given by $T_{D4}= {1\over (2\pi)^4 l_{s}^{5}}$, the gauge coupling constant $g_{YM}$ is identified to be 
\begin{eqnarray}
g_{YM}^{2}={(2\pi)^2 g_{s} l_{s} \over \delta \tau}.
\label{YM}
\end{eqnarray} 

Using (\ref{R}),(\ref{period}), and (\ref{YM}), we have
\begin{eqnarray}
U_{R}^{3} U_{KK} = {1\over 9} \lambda^{2} l_{s}^{4}
\label{RU}
\end{eqnarray} 
where $\lambda \equiv g_{YM}^{2} N_{c}$ is the 't Hooft coupling constant. The above holographic description is reliable under the conditions: $g_{YM} \rightarrow 0$, $N_{c} \rightarrow 0$ and $\lambda$ fixed and large \cite{Kruczenski} so that both loop effects and higher-derivative string corrections can be neglected.

\section{Null Melvin Twist}
In this section we will apply a Null Melvin Twist \cite{NMT1, NMT2}, which is a solution generating method in supergravity, to the geometry (\ref{d4}). It allows us to start from a supergravity solution and construct a new one with different asymptotic geometry. The method contains following steps:\\

1. Boost the type IIA geometry (\ref{d4}) in a translationally invariant direction (we choose $\tau$) by amount $\gamma$.\\

2. $T$-dualize along $\tau$ to obtain a background of type IIB supergravity.\\

3. Twist some one-form dual to another rotatioanl or translational isometry (we choose $d x^{1}$): $d x^{1} \rightarrow d x^{1} + \alpha d\tau$.\\

4. $T$-dualize along $\tau$ to get back a background of type IIA supergravity.\\

5. Boost by $-\gamma$ along $\tau$ to undo the coordinate transformation in step 1.\\

6. Take the scaling limit $\alpha \rightarrow 0$ with ${1\over 2} \alpha e^{\gamma} \equiv \beta$ fixed.\\

After applying this procedure, the resulting geometry is 
\begin{eqnarray}
ds^2&&={\left({U\over U_{R}}\right)^{{3\over 2}} \over \left(1- \beta^2 \left({U_{KK}\over U_{R}}\right)^{3} \right) } \left[ -\left( 1+\beta^2 f(U)  \left({U \over U_{R}}\right)^{3} \right) dt^2  -2 \beta^2 f(U)  \left({U \over U_{R}}\right)^{3} dt d\tau \right. \nonumber \\
 &&\left. + f(U) \left(1- \beta^2 \left({U \over U_{R}}\right)^{3}\right)d\tau^2 + (dx^1)^2 \right] +\left( {U\over U_{R} } \right)^{3\over 2}\left( (dx^{2})^{2} + (dx^{3})^{2} \right)  \nonumber \\
 &&+ \left({U\over U_{R}}\right)^{-{3\over 2}} \left( {dU^2 \over f(U)} + U^2 d \Omega_{4}^{2} \right) \nonumber\\
B_2=&& {-\beta \left({U\over U_{R}}\right)^{3} \over \left(1- \beta^2 \left({U_{KK}\over U_{R}}\right)^{3} \right) }\left( dt+f(U) d\tau \right) \wedge dx^{1},\;\; \;\; e^{\phi}={g_s \over \left(1- \beta^2 \left({U_{KK}\over U_{R}}\right)^{3} \right)^{{1\over2}} }\left( {U\over U_{R}}\right)^{{3\over4}}.
\label{nonrela}
\end{eqnarray} 
The four-form R-R field strength $F_{(4)}$ is not affected by the Null Melvin Twist. Also, the scale dependence of the dilaton remains the same. Instead, a light-like NS-NS two-form is generated in the new solution. To make physical sense, $\beta$ is bounded from above as
\begin{eqnarray}
\beta^2 < {U_{R}^{3}\over U_{KK}^3}={1\over9} {\lambda^2 l_{s}^{4} \over U_{KK}^{4}}.
\end{eqnarray}

By rescaling the coordinates as
\begin{eqnarray} 
t \rightarrow {1 \over \sqrt{1- \beta^2 \left({U_{KK}\over U_{R}}\right)^{3} }} t, \tau \rightarrow {1 \over \sqrt{1- \beta^2 \left({U_{KK} \over U_{R}} \right)^{3} }} \tau, x^1 \rightarrow {1 \over \sqrt{1- \beta^2 \left({U_{KK}\over U_{R}}\right)^{3} }} x^1, \nonumber
\end{eqnarray}
we have
\begin{eqnarray}
ds^2&&=\left({U\over U_{R}}\right)^{{3\over 2}}  \left[ -\left( 1+ \beta^2 f(U)  \left({U \over U_{R}}\right)^{3} \right) dt^2  -2 \beta^2 f(U)  \left({U \over U_{R}}\right)^{3} dt d\tau \right. \nonumber  \\
 &&\left. + f(U) \left(1- \beta^2 \left({U \over U_{R}}\right)^{3}\right)d\tau^2 + \displaystyle\sum_{i=1}^3(dx^i)^2  \right] + \left({U\over U_{R}}\right)^{-{3\over 2}} \left( {dU^2 \over f(U)} + U^2 d \Omega_{4}^{2} \right) \nonumber\\
B_2=&& -\beta \left({U\over U_{R}}\right)^{3} \left( dt+f(U) d\tau \right) \wedge dx^{1},\;\; \;\; e^{\phi}={g_s \over \left(1- \beta^2 \left({U_{KK}\over U_{R}}\right)^{3} \right)^{{1\over2}} }\left( {U\over U_{R}}\right)^{{3\over4}}.
\label{nonrela2}
\end{eqnarray}
The holographic geometry of pure QCD can be obtained from (\ref{nonrela2}) by taking the limit $\beta \rightarrow 0$.

From (\ref{nonrela2}), one can see that the symmetry group of the world-volume is broken from $ISO(1,3)$ to $ISO(3)$ after the specific Null Melvin Twist we performed. The $SO(5)$ rotational symmetry around the transverse coordinates is left untouched. Note that in the third step of Null Melvin Twist, if we chose to twist along a direction transverse to the world-volume of the $D4$ brane, the resulting metric would depend on the angular coordinates of the $S^4$ \cite{Pal, Oz}. We will not consider this case in our paper. 

\section{ Embedding in M-theory}
In the holographic approach, the radial coordinate $U$ corresponds to the energy scale. Since the dilaton is scale dependent as shown in (\ref{nonrela2}), the type IIA geometry is unreliable in the UV. In this section, we lift the geometry (\ref{nonrela2}) to M-theory for a UV-completion.

M-theory compactified on a circle corresponds to type IIA superstring theory. Thus we can consider the type IIA background (\ref{nonrela2}) as the one constructed from eleven-dimensional supergravity by dimensional reduction. More precisely, the eleven-dimensional metric is decomposed into 
\begin{eqnarray}
G_{MN} = e^{-{2\phi \over 3}} \left( \begin{array}{cc}
g_{\mu\nu} +e^{2\phi} C_{\mu} C_{\nu}  & e^{2\phi} C_{\mu}   \\
e^{2\phi} C_{\nu} &   e^{2\phi} \end{array} \right)
\end{eqnarray}
where $C_{\mu}$ is a $U(1)$ R-R gauge potential and all of the fields depend on the ten-dimensional spacetime coordinates only. Therefore, the eleven-dimensional metric is
\begin{eqnarray}
ds_{11}^{2}= G_{MN} dx^{M} dx^{N} = e^{-{2\phi \over 3}} ds_{10}^{2} +e^{{4\phi \over 3}}(dx^{10} + C_{\mu} d^{\mu})^2
\label{M metric}
\end{eqnarray}
where $ds_{10}^2$ is the type IIA metric. For more details on lifting type IIA superstring theory to M-theory, see \cite{Becker}. 

Plugging (\ref{nonrela2}) into the M-theory metric (\ref{M metric}), the resulting $M5$-brane metric is given by 
\begin{eqnarray}
ds_{11}^{2}=&& \left({U\over U_{R}}\right) \left[ -\left( 1+ \beta^2 f(U)  \left({U \over U_{R}}\right)^{3} \right) dt^2  -2  \beta^2 f(U)  \left({U \over U_{R}}\right)^{3} dt d\tau+ f(U) \left(1- \beta^2 \left({U \over U_{R}}\right)^{3}\right)d\tau^2 \right. \nonumber  \\
&&  \left. + \displaystyle\sum_{i=1}^3(dx^i)^2 +(dx^{10})^2+  {dU^2 \over \left( {U\over U_{R}} \right)^3- \left( {U_{KK} \over U_{R}} \right)^3} + U^2 d \Omega_{4}^{2}  \right] . 
\label{11d}
\end{eqnarray}
To obtain the above result, we have rescaled the coordinates to absorb the dimensionless coupling constant. 

From now on, we will investigate the six-dimensional metric obtained from (\ref{11d}) by compactifying on the $S^4$ and the $x^{10}$ circle, and consider it to be the geometrical realization of a nonrelativistic field theory. The truncated six-dimensional metric reads
\begin{eqnarray}
ds^{2}= \left({U\over U_{R}}\right)&& \left[  -\left( 1+ \beta^2 f(U)  \left({U \over U_{R}}\right)^{3} \right) dt^2  -2 \beta^2 f(U)  \left({U \over U_{R}}\right)^{3} dt d\tau+ f(U) \left(1- \beta^2 \left({U \over U_{R}}\right)^{3}\right)d\tau^2  \right.  \nonumber \\
 &&  \left. + \displaystyle\sum_{i=1}^3(dx^i)^2 +  {dU^2 \over \left( {U\over U_{R}} \right)^3- \left( {U_{KK} \over U_{R}} \right)^3} \right]. 
\label{6d}
\end{eqnarray} 

To summarize, from the above discussion one can see that the pure QCD dual can arise from the M-theory background (\ref{11d}) by dimensional reduction in the limit $\beta\rightarrow 0$. On the other hand, one can also obtain from the same metric a lower-dimensional background which geometrically realizes the symmetries of a nonrelativistic field theory.

\section{Z=4 dual}
First we would like to show that the metric (\ref{6d}) is asymptotically conformal to a $pp$-wave spacetime. That is, we will consider the region where $U \gg U_{KK}$. It is easy to see this by changing coordinates $u\equiv {1\over \sqrt{2}} (\tau+t)$, $v\equiv {1\over \sqrt{2}} (\tau-t)$, and $ {U\over U_{R}}\equiv ({2U_{R}\over r})^2$. The result in coordinates $\{ u, v, r , \vec{x}\}$ is
\begin{eqnarray} 
ds^{2}= {(2U_{R})^2 \over r^2} \left( -2\beta (2U_{R})^6 {1\over r^6} du^2 +2dudv + \displaystyle\sum_{i=1}^3(dx^i)^2+ dr^2 \right). 
\label{z4}
\end{eqnarray} 
In this coordinate system it is manifest the above metric is conformal to a $pp$-wave spacetime, and is usually denoted by $Sch_{6}^{4}$, where $6$ counts the dimension of spacetime and $4$ is the dynamical exponent. The isometries include time translation, translations and rotations in three spatial coordinates $x^i$, a dilatation symmetry 
\begin{eqnarray} 
( u,v,r,\vec{x}) \rightarrow (s^4 u, s^{-2} v, s r, s\vec{x} ) 
\label{scaling}
\end{eqnarray} 
and the Galilean boosts
\begin{eqnarray} 
\left( \begin{array}{c}
v \\
\vec{x}  \\
u  \end{array} \right)\rightarrow \left( \begin{array}{ccc}
1 & -\vec{k} & -{k^2 \over 2} \\
0 & 1_{3\times3} & -\vec{k} \\
0 & 0 & 1 \end{array} \right)\left( \begin{array}{c}
v \\
\vec{x}  \\
u  \end{array} \right).
\label{boost}
\end{eqnarray} 
In the prescription of gauge/gravity duality, the metric (\ref{z4}) is interpreted as the geometrical realization of a four-dimensional nonrelativistic field theory which exhibits the above symmetries with dynamical exponent $z=4$. 

Note that in the coordinate system $\{ u, v, r , \vec{x}\}$, the coordinate $r$ corresponds to the RG length scale in the holographic picture. Thus the original bulk coordinte $U$ should scale as energy square $U\sim E^2$. The interpretation of the $v$-momentum as the particle number requires the coordinate $v$ to be periodic. This is guaranteed by the compactifiction of $\tau$. 

Let us now go back to the metric (\ref{6d}). Like its QCD cousin, the horizon is at $U=U_{KK}$. From (\ref{6d}) it is clear that the Killing vector, which is defined to be a non-affinely parametrized geodesic on the Killing horizon, is $({\partial \over \partial \tau})^{a}$. Thus it is easy to show that the period of $\tau$ is given by 
\begin{eqnarray}
\Delta \tau = \lim _{U\rightarrow U_{KK}}  {4\pi\over \sqrt{g^{UU} g^{\tau\tau} \partial_{U} (g_{\tau\tau})^2}}={4\pi \over 3} \left(1-\beta^2 \left( {U_{KK} \over U_{R}} \right)^3  \right)^{-{1\over2}} \left( {U_{R}^{3}\over U_{KK} } \right)^{{1\over2}}.
\end{eqnarray}
Up to the factor $\left(1-\beta^2 \left( {U_{KK} \over U_{R}} \right)^3  \right)^{-{1\over2}}$, $\Delta \tau$ is the same as the period of $\tau$ in the holographic QCD model (\ref{period}), and the origin of this factor comes from our rescaling of coordinates made before (\ref{nonrela2}).

The metric (\ref{6d}) is nonsingular. In fact, the Ricci scalar $R$ is constant and takes the value 
\begin{eqnarray}
R={3\over 2U_{R}^{2}}\left( {3\over 1- \beta^2 {U_{KK}^{3} \over U_{R}^3}} - 8\right).
\label{Ricci}
\end{eqnarray}
Note that the Ricci scalar is constant only when the dimensionality of the $\vec{x}$ equals 3 or 4.

Recall that the metric (\ref{6d}) becomes $Sch_{6}^{4}$ as the non-extremal parameter $U_{KK}\rightarrow 0$. In the limit $U_{KK}\rightarrow 0$ and $\beta \rightarrow 0$, (\ref{6d}) is sent to the metric of $AdS_{6}$. From (\ref{Ricci}) it it clear that both $Sch_{6}^{4}$ and $AdS_{6}$ obtained from (\ref{6d}) have the same Ricci scalar $R=-{15\over 2 U_{R}^2}$.

Similar to the construction of the finite temperature QCD dual described in \cite{Johnson}, one can build the finite temperature version of the nonrelativistic model (\ref{6d}). To do this, we need to compactify the time direction on a circle. For low temperature, one simply consider the Euclidean version of (\ref{6d}) and interpret the period of $t$ as the inverse temperature. Foe high temperature, the metric is given by the Euclidean version of (\ref{6d}) under interchange of $t$ and $\tau$. Explicitly, the high temperature metric takes the form 
\begin{eqnarray}
ds^{2}= \left({U\over U_{R}}\right)&& \left[  \left( 1+ \beta^2 f(U)  \left({U \over U_{R}}\right)^{3} \right) d\tau^2  -2 \beta^2 f(U)  \left({U \over U_{R}}\right)^{3} dt d\tau+ f(U) \left(1- \beta^2 \left({U \over U_{R}}\right)^{3}\right)dt^2  \right.  \nonumber \\
 &&  \left. + \displaystyle\sum_{i=1}^3(dx^i)^2 +  {dU^2 \over \left( {U\over U_{R}} \right)^3- \left( {U_{KK} \over U_{R}} \right)^3} \right]. 
\label{HT}
\end{eqnarray} 

Parallel to the calculation for the period of $\tau$ in the zero temperature background, the temperature of the background (\ref{HT}) is given by 
\begin{eqnarray}
T = {3\over 4\pi } \left(1-\beta^2 \left( {U_{KK} \over U_{R}} \right)^3  \right)^{{1\over2}} \left( {U_{KK}\over U_{R}^{3} } \right)^{{1\over2}}.
\label{T}
\end{eqnarray} 

\section{Discussion}
One significant feature behind the holography that we would like to emphasize in this paper is the uniqueness and universality of gravity. It allows us to relate theories with different dynamical exponents. In this paper, we have focused on pure QCD as an example and showed by construction that both the pure QCD dual and the nonrelativistic field theory geometrically realized by a deformed $Sch_{6}^{4}$ can be embedded in the same M-theory background. This is the main goal of this paper. 

We have discussed aspects of the deformed $Sch_{6}^{4}$ constructed from QCD. There are several directions for further research:

$\bullet$ With the constructed geometry, one can compute the two point function of the stress-tensor and the shear viscosity following the recipes of \cite{Buchel,Son2}. 

$\bullet$We have constructed only the geometry. It would be interesting to concoct a six-dimensional gravitational model whose solution is given by the deformed $Sch_{6}^{4}$ (\ref{6d}).

$\bullet$ By adding "probe fields" in the finite temperature backgrounds we found, it should be stimulating to investigate the transition between hot and cold backgrounds and relate it to the confinement/deconfinement in QCD dual.

$\bullet$ As was point out in \cite{Balasubramanian} by considering a scalar in the $Sch_{d}^{z}$ background, theories with dynamical exponent $z>2$ have no scaling solution near the boundary. Thus a better understanding of the operator-field correspondence is desirable to properly extend the holographic dictionary to these systems.


\begin{acknowledgments}
The research of F.W. was supported in part by the National Nature Science Foundation of China under grants No. 10805024 and No. 10965003, and the project of Chinese Ministry of Education under grant No. 208072.
\end{acknowledgments}




\begin{thebibliography}{99}                                                                                               
\bibitem {Maldacena}J. Maldacena, Adv. Theor. Math. Phys. \textbf{2}, 231 (1998) [ Int. J. Theo. Phys. \textbf{38}, 1113 (1999)].

\bibitem {Witten1}A. Witten, Adv. Theor. Math. Phys. \textbf{2}, 253 (1998).

\bibitem {Gubser}S. S. Gubser, I. R. Klebanov and A. M. Polyakov, Phys. Lett. B \textbf{428}, 105 (1998).

\bibitem {Sakai}T. Sakai and S. Sugimoto, Prog. Theor. Phys. \textbf{113}, 843 (2005); Prog. Theor. Phys. \textbf{114}, 1083 (2006).

\bibitem {Witten2}A. Witten, Adv. Theor. Math. Phys. \textbf{2}, 505 (1998).

\bibitem {fermions}K. M. O'Hara et. al., Science \textbf{298}, 2179 (2002); C. A. Regal, M. Greiner and D. S. Jin, Phys. Rev. Lett. \textbf{92}, 040403 (2004); M. Bartenstein et. al., Phys. Rev. Lett. \textbf{92}, 120401 (2004); M. Zwierlein et. al., Phys. Rev. Lett. \textbf{92}, 120403 (2004); J. Kinast et. al., Phys. Rev. Lett. \textbf{92}, 150402 (2004); T. Bourdel et. al., Phys. Rev. Lett. \textbf{93}, 050401 (2004).

\bibitem {Son} D. T. Son, Phys. Rev. D \textbf{78}, 046003 (2008).

\bibitem {Balasubramanian}K. Balasubramanian and J. McGreevy, Phys. Rev. Lett. \textbf{101}, 061601 (2008).

\bibitem {Kachru}S. Kachru, X. Liu and M. Mulligan, Phys. Rev. D \textbf{78}, 106005 (2008).

\bibitem {others}J. L. F. Barbon and C. A. Fuertes, JHEP \textbf{0809}, 030 (2008); W. D. Goldberger, arXiv:0806.2867 [hep-th];J. Maldacena, D. Martelli and Y. Tachikawa, JHEP \textbf{0810}, 072 (2008); A. Adams, K. Balasubramanian and J. McGreevy, JHEP \textbf{0811}, 059 (2008); P. Kovtun and D. Nickel, Phys. Rev. Lett. \textbf{102}, 011602 (2009); C. Duval, M. Hassaine and P. A. Horvathy, Annals. Phys. \textbf{324}, 1158 (2009); S.-S. Lee, Phys. Rev. D \textbf{79}, 086006 (2009).

\bibitem {NMT1}M. Alishahiha and O. J. Ganor, JHEP \textbf{0303}, 006 (2003).

\bibitem {NMT2}E. G. Gimon, A. Hashimoto, V. E. Hubeny, O. Lunin and M. Rangamani, JHEP \textbf{0308}, 035 (2003).

\bibitem {Pal}S. Pal, arXiv:0808.3042 [hep-th].

\bibitem {Herzog}C. P. Herzog, M. Rangamani and S. F. Ross, JHEP \textbf{0811}, 080 (2008).

\bibitem {Oz}L. Mazzucato, Y. Oz and S. Theisen, JHEP \textbf{0904}, 073 (2009).

\bibitem {Kruczenski}M. Kruczenski, D. Mateos, R. C. Myers and D. J. Winters, JHEP \textbf{0405}, 041 (2004).

\bibitem {Becker}K. Becker, M. Becker and J. H. Schwarz, String Theory and M-theory: A Modern Introduction, Cambridge University Press (2007).

\bibitem {Johnson}C. V. Johnson and A. Kundu, JHEP \textbf{0812}, 053 (2008).

\bibitem {Buchel}A. Buchel and J. T. Liu, Phys. Rev. Lett. \textbf{93}, 090602 (2004).

\bibitem {Son2}P. Kovtun, D. T. Son and A. O. Starinets, Phys. Rev. Lett. \textbf{94}, 111601 (2005).

\end{thebibliography}
\end{document}